\documentclass{article}
\usepackage{spconf,amsmath,graphicx}
\usepackage{subfigure}
\usepackage{multirow}
\usepackage[urlcolor=blue]{hyperref}
\usepackage{booktabs}
\usepackage{amsfonts}
\usepackage{float}
\usepackage{amsthm,amsmath,amssymb}
\usepackage{bbding}
\usepackage{color}

\title{StyleSpeech: Self-supervised Style Enhancing with VQ-VAE-based Pre-training for Expressive Audiobook Speech Synthesis}
\makeatletter
\def\name#1{\gdef\@name{#1\\}}
\makeatother
\name{
    \em{Xueyuan Chen$^{1,2,\dagger}$\thanks{$\dagger$ Work conducted when the first author was intern at Microsoft.}, Xi Wang$^3$, Shaofei Zhang$^3$, Lei He$^3$, Zhiyong Wu$^{1,2,*}$\thanks{* Corresponding author.}, {Xixin Wu}$^{1}$,{Helen Meng}$^{1,2}$}}

\address{
    $^1$ Department of Systems Engineering and Engineering Management, \\
         The Chinese University of Hong Kong, Hong Kong SAR, China\\
    $^2$ Tsinghua-CUHK Joint Research Center for Media Sciences, Technologies and Systems, \\
    Shenzhen International Graduate School, Tsinghua University, Shenzhen, China\\
    $^3$ Microsoft, Beijing, China\\
    \small{ 
        \{xychen,zywu,wuxx, hmmeng\}@se.cuhk.edu.hk, \{xwang, shazh, helei\}@microsoft.com
    }}

\begin{document}
\ninept
\maketitle
\begin{abstract}
\vspace{-3pt}
The expressive quality of synthesized speech for audiobooks is limited by generalized model architecture and unbalanced style distribution in the training data.
To address these issues, in this paper, we propose a self-supervised style enhancing method with VQ-VAE-based pre-training for expressive audiobook speech synthesis. 
Firstly, a text style encoder is pre-trained with a large amount of unlabeled text-only data. 
Secondly, a spectrogram style extractor based on VQ-VAE is pre-trained in a self-supervised manner, with plenty of audio data that covers complex style variations. 
Then a novel architecture with two encoder-decoder paths is specially designed to model the pronunciation and high-level style expressiveness respectively, with the guidance of the style extractor.
Both objective and subjective evaluations demonstrate that our proposed method can effectively improve the naturalness and expressiveness of the synthesized speech in audiobook synthesis especially for the role and out-of-domain scenarios.\footnote[1]{\href{https://Chenxuey20.github.io/StyleSpeech}{https://Chenxuey20.github.io/StyleSpeech}}

\end{abstract}

\begin{keywords}
expressive speech synthesis, self-supervised style enhancing, VQ-VAE, pre-training
\end{keywords}

\vspace{-5pt}
\section{introduction}
\vspace{-5pt}

Recent text-to-speech (TTS) models, e.g., Tacotron 2 \cite{shen2018natural}, TransformerTTS \cite{li2019neural}, FastSpeech 2 \cite{renfastspeech}, have been developed with the capability to generate high-quality speech with a neutral speaking style.
However, limited expressiveness 
persists as one of the major gaps between synthesized speech and real human speech,
which draws growing attention to expressive speech synthesis studies \cite{wang2018style, chen2022character, skerry2018towards, chen2022hilvoice}.
Synthesizing long-form expressive datasets, e.g., audiobooks, is still a challenging task,
since wide-ranging voice characteristics 
tend to collapse into an averaged prosodic style.

There are a lot of works focusing on audiobook speech synthesis \cite{charfuelan2013expressive, stanton2018predicting, sini2018synpaflex}.
Recently, \cite{xu2021improving} proposes to use the neighbor sentences to improve the prosody generation.
To make better use of contextual information, a hierarchical context encoder that considers 
adjacent sentences
with a fixed-size sliding window is used to predict a global style representation directly from text \cite{lei2022towards}.
Besides, \cite{wu2022discourse} tries to consider as much information as possible (e.g., BERT embeddings, text embeddings and sentence ID) to improve style prediction. 
On top of these, a multi-scale hierarchical context encoder is proposed to predict both global-scale and local-scale style embeddings from context in a hierarchical structure \cite{chen2022unsupervised}.
All these existing works mainly focus on how to use the semantic information of contextual text to predict the expressiveness through an additional style encoder module.
Too much information (phoneme, timbre, style, etc.) is simply mixed in the encoder part, leading to challenges for mel-spectrogram decoder.
In addition, another serious problem for audiobook synthesis is the unbalanced  style distribution in audiobook dataset.
Most sentences are relatively plain narration voices, and only a small part is role voices with rich style variations,
which brings a great challenge to modeling of style and expressiveness representation with limited audiobook training data, especially for role and out-of-domain scenarios.

To solve the above-mentioned poor expressiveness problem in audiobook speech synthesis caused by generalized model architecture and unbalanced style distribution in the training data,
this paper proposes a self-supervised style enhancing method with VQ-VAE-based pre-training for expressive audiobook synthesis.
Firstly, a text style encoder is pre-trained with the help of a large amount of easily obtained unlabeled text-only data.
Secondly, a spectrogram style extractor based on VQ-VAE is pre-trained using plenty of audio data that covers multiple expressive scenarios in other domains.
On top of these, a special model architecture is designed with two encoder-decoder paths with the guidance of style extractor.
To summarize, the main contributions of this paper are:
\begin{itemize}
\item We propose a VQ-VAE-based style extractor to model a better style representation latent space and relieve the unbalanced style distribution issues,
which is pre-trained by plenty of easily obtained audio data that can cover complex style variations in a self-supervised manner.

\item We design a novel TTS architecture with two encoder-decoder paths to model the pronunciation and 
high-level style expressiveness respectively, 
so as to enrich the expressive variation of synthesized speech in complex scenarios by strengthening both the encoder and decoder of TTS model.

\item Both objective and subjective experimental results show that our proposed style enhancing approach achieves 
an effective
improvement in terms of speech naturalness and expressiveness
especially for the role and out-of-domain scenarios.
\end{itemize}

\begin{figure*}[ht]
\centering
\includegraphics[width=2.05\columnwidth]{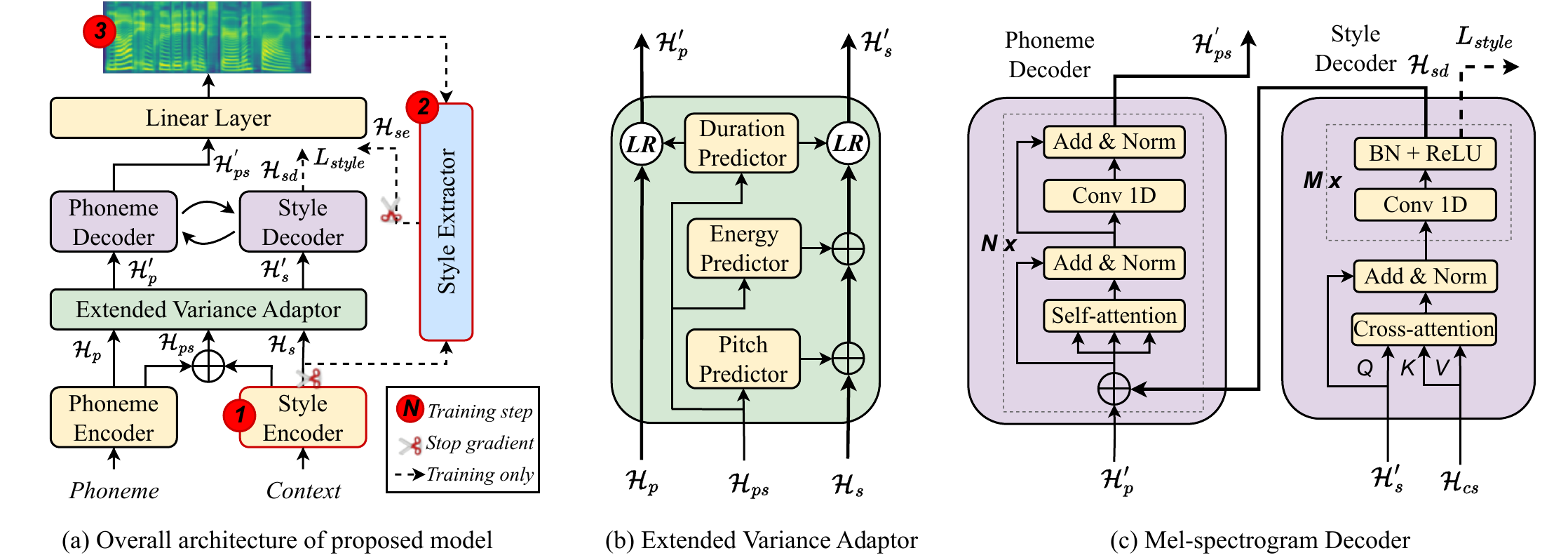}
\vspace{-10pt}
\caption{Proposed model structure, where (a) shows the overall architecture with two encoder-decoder paths, (b) shows details of Extended Variance Adaptor and (c) shows Mel-spectrogram Decoder containing Phoneme Decoder and Style Decoder with interaction.}
\vspace{-10pt}
\label{fig:model_structure}
\end{figure*}

\vspace{-8pt}
\section{Related work}
\vspace{-5pt}
Our work is related to Context-aware Augmented Deep Embedded Clustering (CADEC) \cite{wu2022self} and Vector Quantized-Variational AutoEncoder (VQ-VAE) \cite{van2017neural}.
\vspace{-8pt}
\subsection{CADEC}
CADEC is a 
two-stage style learning approach from abundant unlabeled plain text in a self-supervised manner.
Firstly, it uses contrastive learning \cite{chen2020simple} to pre-train style embedding to distinguish similar and dissimilar utterances. 
To this end, a similar utterance is created by replacing an emotional word, determined by an emotion lexicon, with a similar one, while the other utterances in the randomly sampled minibatch are treated as dissimilar utterances. 
Secondly, the training samples in style embedding space are clustered by minimizing deep clustering loss \cite{zhang2021supporting}, reconstruction loss and contrastive loss together. 
Compared with BERT \cite{kenton2019bert}, CADEC style embedding is more effective in learning styles other than content.
\vspace{-8pt}

\subsection{VQ-VAE}
VQ-VAE is a powerful representation learning framework that can make effective use of the latent space.
It 
combines VAE framework with discrete latent representations through a 
parameterisation of the posterior distribution of (discrete) latents given an observation.
It can successfully model important features that usually span many dimensions in data space (e.g., objects span many pixels in images, phonemes in speech, the message in a text fragment, etc.) as opposed to focusing or spending capacity on noise and imperceptible details which are often local.
Many extension models have been proposed, leading to high performance in various tasks,
e.g., prosody learning \cite{zhao2020improved} and speaker diarization \cite{williams2021learning}. 
\vspace{-8pt}

\section{methodology}
\vspace{-5pt}
The overall architecture of our proposed model is illustrated in Fig. \ref{fig:model_structure} (a).
It mainly consists of two encoder-decoder paths with interaction.
The first and primary one is the fine-grained phoneme path,
while the second one is the coarse-grained style path.

\vspace{-8pt}
\subsection{Phoneme encoder-decoder path}
The phoneme encoder-decoder path mainly focuses on the 
pronunciation based on FastSpeech 2 \cite{renfastspeech}.
Both the phoneme encoder and phoneme decoder consist of several feed-forward Transformer (FFT) blocks, which are a stack of self-attention layer and 1D-convolution with residual connection and layer normalization. 
As shown in Fig. \ref{fig:model_structure} (b), the phoneme hidden embedding $\mathcal{H}_p$ is repeated to frame-level phoneme representation $\mathcal{H}_p^{'}$ by length regulator (LR) in the extended variance adaptor.
And it is worth noting that only 
$\mathcal{H}_p^{'}$ is further fed to the phoneme decoder in this path, not together with the pitch and energy,
which is different from FastSpeech 2.

\vspace{-8pt}
\subsection{Style encoder-decoder path}
The style encoder-decoder path 
focuses on the style modeling 
of synthesized speech.
Specifically, a text style encoder and  a spectrogram style extractor are designed  and pre-trained to learn 
the style-related representations from 
contextual text
and mel-spectrogram respectively,
with a huge amount of 
unlabeled data.
A style decoder is further adopted to make a better fusion of the explicit style features and implicit style representations in the decoding stage.

\vspace{-8pt}
\subsubsection{Style Encoder}
We adopt the CADEC encoder \cite{wu2022self} as our style encoder.
It employs a pre-trained BERT \cite{kenton2019bert} as backbone to extract semantic features, and an 
emotion lexicon \cite{buechel2020learning} to extract emotion features.
By contrastive learning with data augmentation and deep embedded clustering with an autoencoder structure,
it can be trained with abundant unlabeled pain text 
and extract a more style-related representation from context.
Finally, by accepting the contextual text $C$, 
CADEC encoder can output a global style representation:
\begin{equation}
    \begin{split}
        \small
        \mathcal{H}_s = CADEC(C_0)
    \end{split}
\end{equation}
\begin{equation}
    \begin{split}
        \small
        \mathcal{H}_{cs} = Concat[CADEC(C_i),i=-k,...,k]
    \end{split}
\end{equation}
where $Concat[.]$ is the concatenation operation, $\mathcal{H}_s$ and $\mathcal{H}_{cs}$ are the hidden style embeddings for the contextual text $C_0$ of the current utterance and for the $2k+1$ neighbor utterances respectively.

\vspace{-8pt}
\subsubsection{Extended Variance Adaptor}
Based on FastSpeech 2, the extended variance adaptor is designed to explicitly model the style-related features e.g, duration, pitch and energy.
As shown in Fig. \ref{fig:model_structure} (b), 
the phoneme encoder output $\mathcal{H}_p$ and style encoder output $\mathcal{H}_s$ are added together (denoted as $\mathcal{H}_{ps}$) to fed into the pitch predictor, energy predictor and duration predictor respectively to predict the phoneme-level explicit style features,
which are related to both the phoneme and style.
Furthermore, the predicted pitch and energy together with the implicit style embedding $\mathcal{H}_s$ are repeated to become the frame-level style embedding $\mathcal{H}_{s}^{'}$ by the length regulator.

\vspace{-8pt}
\subsubsection{Style Extractor}
\begin{figure}[tb]
	\centering
	\includegraphics[width=0.8\columnwidth]{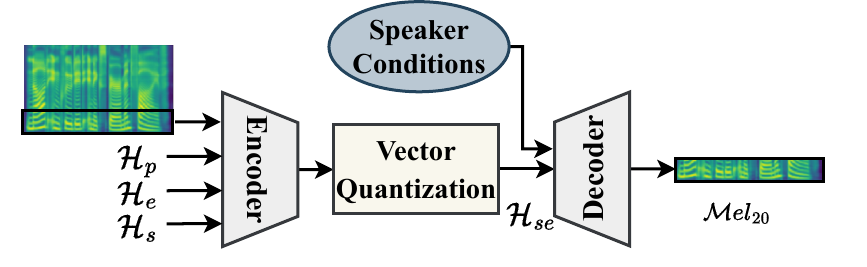}
        \vspace{-10pt}
	\caption{Style Extractor based on VQ-VAE.}
        \vspace{-15pt}
	\label{fig:style_extractor}
\end{figure}

As shown in Fig. \ref{fig:style_extractor}, we adopt VQ-VAE \cite{van2017neural}  as our style extractor to extract a 
style-related latent representation from mel-spectrogram with a large amount of unlabeled audio data.
Specifically, the encoder consists of two 2D-convolution layers with batch normalization (BN) and ReLU activation, followed by several ResBlock \cite{he2016deep} layers,
while the decoder adopts a symmetrical structure with the encoder.
Besides, the one-hot speaker embedding conditions are also fed to decoder to remove the influence of timbres.

Only the low-frequency band of the mel-spectrogram $\mathcal{M}el_{20}$ (first 20 bins in each 
frame) is taken as input,
as it is considered to contain almost complete style and much less content information compared with the full band.
Besides, in order to further guide the model to extract style-related latent representations, we also use the frame-level pitch $\mathcal{H}_p$, frame-level energy $\mathcal{H}_e$ and text features $\mathcal{H}_s$ as additional inputs.
Finally, a discrete style-related representation $\mathcal{H}_{se}$ can be extracted from the vector quantization layer output of well-pretrained style extractor, which can be described as follows: 
\begin{equation}
    \begin{split}
        \small
        \mathcal{H}_{se} = VQVAE(\mathcal{M}el_{20},\mathcal{H}_p,\mathcal{H}_e,\mathcal{H}_s)
    \end{split}
\end{equation}

\vspace{-10pt}
\subsubsection{Style Decoder}
The style decoder is designed to further integrate the explicit style features (pitch, energy) and implicit style embeddings
in the decoding stage.
As shown in Fig. \ref{fig:model_structure} (c),
in order to make the style transitions among contextual sentences more natural and smooth,
a cross-attention module followed by residual connection is firstly adopted to consider the hierarchical context.
Here, the frame-level style embedding $\mathcal{H}_s^{'}$ of current utterance is the query,
while the hierarchical context style embedding $\mathcal{H}_{ce}$ from style encoder is the key and value.
After that, several 1D-convolution layers with batch normalization and ReLU activation are further used to learn 
a style-related representation
and finally output the style embedding $\mathcal{H}_{sd}$.

\vspace{-8pt}
\subsection{Interaction between phoneme and style paths}
Existing expressive speech synthesis works mainly simply introduce style information into the TTS encoder part,
leading to challenges to the mel-spectrogram decoder.
As shown in Fig. \ref{fig:model_structure} (c),
we make the feature interaction between phoneme and style paths not only in the encoder part but also in the mel-spectrogram decoder part.
Specifically, the output embedding $\mathcal{H}_{sd}$ of style decoder is fed into each FFT block of phoneme decoder as an additional style input in order to fully integrate the style and pronunciation information. 
After that, the 
well-mixed output embedding $\mathcal{H}_{ps}^{'}$ of mel-spectrogram decoder based on the two encoder-decoder paths is finally 
fed into the post linear layer to reconstruct the mel-spectrogram.

\vspace{-8pt}
\subsection{Training strategy and inference procedure}
As shown in Fig. \ref{fig:model_structure}(a), our proposed model is trained in three stages.

i) In the first stage, the style encoder is pre-trained with a large amount of text data.
Training details are similar to \cite{wu2022self}.

ii) In the second stage, the style extractor is trained with a large amount of audio data.
Consistent with the original VQ-VAE, the total training loss consists of a reconstruction loss for reconstructing the mel-spectrogram, a vector quantisation loss for updating the dictionary  and a commitment loss for making sure the encoder commits to an embedding.

iii) In the third stage, the TTS model is trained with audiobook data.
The model parameters of style encoder and style extractor are frozen without gradient update.
An additional style loss $\mathcal{L}_{style}$ is adopted to the style encoder-decoder path to give a guidance from the pre-trained style extractor.
\vspace{-5pt}
\begin{equation}
    \begin{split}
        \small
        \mathcal{L}_{style} = MSE(\mathcal{H}_{sd}, \mathcal{H}_{se})
    \end{split}
\vspace{-10pt}
\end{equation}
where $MSE$ is the mean square error (MSE) loss, $\mathcal{H}_{sd}$ and $\mathcal{H}_{se}$ are the outputs of style decoder and style extractor respectively.
The total loss is as follows:
\vspace{-5pt}
\begin{equation}
\vspace{-4pt}
    \begin{split}
        \small
        \mathcal{L}_{total} = \mathcal{L}_{tts} + \alpha \mathcal{L}_{style}
    \end{split}
\end{equation}
where $\mathcal{L}_{tts}$ is the TTS loss consistent with FastSpeech 2. 

During inference, the style extractor is abandoned (as shown by the dotted line in Fig. \ref{fig:model_structure} (a)).
By accepting phoneme and context input, the model can synthesize speech with more expressive styles.

\vspace{-5pt}
\section{Experiments}
\vspace{-5pt}

\subsection{Datasets and system settings}
We use 3 types of internal Mandarin datasets to train style encoder, style extractor and TTS model respectively.
The style encoder is trained with a large plain text dataset, containing 7.5M audiobook sentences.
The style extractor is trained with a large multi-speaker audio corpus,
which contains around 400 hours of audios with corresponding text and covers a wealth of application scenarios and style variations.
We use an audiobook corpus to train the TTS model.
It has around 
30-hour speech data
with context and is cut into 30,000 audio clips, of which 1000 clips are used for validation and 500 clips for test, and the rest for training.
Besides, another small audiobook dataset covering several different categories is further used to evaluate the out-of-domain performance. 
Details are shown in Table \ref{tab:dataset}.
\vspace{-20pt}

\begin{table}[H]
\renewcommand{\arraystretch}{0.9}
\caption{Datasets of different training stages.}
\label{tab:dataset}
\centering
\begin{tabular}{lccc}
\toprule
Stage  & Style Encoder  & Style Extractor  & TTS model  \\ 
\hline
Type  & Text  & Audio    & Audiobook  \\
Size  & 7.5M  & 400 hours    &  30 hours \\
\bottomrule
\end{tabular}
\vspace{-9pt}
\end{table}

\begin{table*}[!htp]
\caption{Subjective and objective evaluation results for different models.}
\label{tab:objective evalution}
\centering
\begin{tabular}{lccccccc}
\toprule
\multirow{2}{*}{\textbf{Model}}  & \textbf{MOS} & \textbf{Style MOS}    & \textbf{Paragraph CMOS}     & \textbf{F0} & \textbf{Energy} & \textbf{Duration} & \multirow{2}{*}{\textbf{MCD}}   \\ 
& \textbf{(out-of-domain)}       & \textbf{(out-of-domain)}  & \textbf{(out-of-domain)}   & \textbf{RMSE} & \textbf{RMSE} & \textbf{MSE} &   \\ 
\hline
Ground Truth  & 4.22 $\pm$ 0.14  & 4.17 $\pm$ 0.11    & -   & -     & -       & -       & - \\
FastSpeech 2  & 4.00 $\pm$ 0.10  & 4.09 $\pm$ 0.13    & -0.161   & 58.930  & 10.848       & 0.0636       & 5.969 \\
FS2-CADEC       & 3.97 $\pm$ 0.10  & 4.08 $\pm$ 0.10    & -0.022   & 58.865  & 10.788       & 0.0629       & 5.951 \\
Proposed      & \textbf{4.08 $\pm$ 0.09} & \textbf{4.17 $\pm$ 0.08}     & \textbf{0}    & \textbf{57.271}  & \textbf{10.697}     & \textbf{0.0617}       & \textbf{5.937} \\
\bottomrule
\end{tabular}
\vspace{-10pt}
\end{table*}

For feature extraction, we transform the raw waveforms into 80-dim mel-spectrograms with sampling rate 16kHz, frame size 1200 and hop size 240.
The context of current sentence is made up of its two past sentences, two future ones and itself.
The codebook size of style extractor is set to 512 and the style loss coefficient $\alpha$ is 1.
All the trainings are conducted with a batch size of 16 on a NVIDIA V100 GPU.
The Adam optimizer is adopted with $\beta_1=0.9, \beta_2=0.98$. 
In addition, a well-trained HiFi-GAN \cite{kong2020hifi} is used as the vocoder to generate waveform.
Two FastSpeech 2 based methods are implemented for comparison as follows:

\begin{itemize}
\item \textbf{FastSpeech 2}: Original FastSpeech 2 \cite{renfastspeech} is implemented as the first baseline method.

\item \textbf{FS2-CADEC}: Inspired by \cite{wu2022self}, we set an end-to-end TTS model by combining CADEC encoder with FastSpeech 2.

\end{itemize}

\vspace{-10pt}
\subsection{Subjective comparison for different systems}
Mean Opinion Score (MOS) is first conducted in terms of the comprehensive performance of synthesized speech including sound quality,
naturalness, expressiveness, etc,
to ensure that all baseline systems are well reproduced.
Furthermore, style MOS is used to only focus on the style expressiveness of synthesized speech,
and paragraph Comparative MOS (CMOS) is used to evaluate the style transition among sentences within a paragraph.
All the tests are conducted on Microsoft UHRS crowdsourcing platform.
As our ultimate goal is to synthesize any other given audiobook,
we mainly focus on the out-of-domain role performance.
50 single sentences and 20 short paragraphs are randomly selected in an out-of-domain set.
Each audio is judged by at least 10 native speakers.

As shown in Table \ref{tab:objective evalution}, our proposed approach achieves the best MOS of 4.08 and best Style MOS of 4.17 compared with the baseline methods.
Specially, our proposed approach achieves comparable results to the ground truth recording on Style MOS.
In the paragraph-level comparison, our proposed approach also achieves the best CMOS performance.
These results demonstrate the effectiveness of our proposed methods 
especially on the role style expressiveness in out-of-domain scenarios.

\vspace{-8pt}
\subsection{Objection comparison for different systems}
For the objective evaluation of synthesized speech,
we employ the root mean square error (RMSE) of pitch and energy, the mean square error (MSE) of duration and mel cepstral distortion (MCD) as the objective evaluation metrics.

As shown in Table \ref{tab:objective evalution}, our proposed model achieves 57.271 for F0 RMSE, 10.697 for Energy RMSE, 0.0617 for duration MSE and 5.937 for MCD,
which outperforms all the baselines on all metrics.
These
results indicate that our proposed model can predict more accurate explicit style features, e.g., duration, pitch and energy, and reconstruct more preserved mel-spectrograms, than baselines.

\vspace{-8pt}
\subsection{Analysis for the pre-training strategy}
To further verify whether the pre-training strategy with plenty of audio data is helpful for the style representation latent space modeling of the unbalanced audiobook data,
we also train a style extractor with only audiobook dataset for comparison.
We extract a few role style embeddings with different style categories in the audiobook dataset by the above-mentioned two well-trained style extractors and make a t-SNE visualization respectively.

\begin{figure}[htb]
	\centering
	\includegraphics[width=0.85\columnwidth]{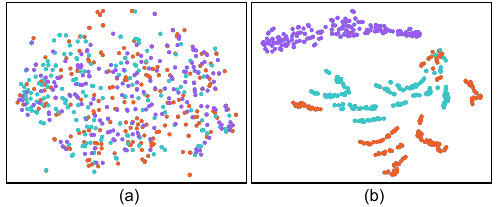}
        \vspace{-10pt}
	\caption{Visualization of style embedding space trained with different datasets. Each color indicates a ground truth style category. (a) represents style embeddings while training with only audiobook dataset. (b) represents style embeddings while training with large dataset.}
        \vspace{-10pt}
	\label{fig:tsne}
\end{figure}

Fig. \ref{fig:tsne} (a) shows the extracted style embeddings when only audiobook dataset participates in training,
and Fig. \ref{fig:tsne} (b) shows the extracted style embeddings when we use the large dataset to train the style extractor.
Obviously, compared to Fig. \ref{fig:tsne} (a), there is a better cohesion and distribution differences among different styles in Fig. \ref{fig:tsne} (b). 
Note that the style in audiobook dataset is too complex to be divided into several categories,
and there may be several slightly different distribution forms even within the same style category by manual annotation.
The results show that it's difficult to model the style representation latent space with only unbalanced and limited audiobook data,
and our proposed pre-training strategy with a large amount of audio data that covers multiple expressive scenarios in other domains is necessary and beneficial for the style latent space modeling of audiobook speech synthesis.

\vspace{-8pt}
\subsection{Ablation study for the model architecture}

To further investigate the influence of several main modules in our proposed model,
we have tried three other settings: 
i) \textbf{Proposed w/o Style Encoder}:
The style encoder is removed, only the output $\mathcal{H}_p$ of phoneme encoder is fed to the extended variance adaptor for both the two paths.
ii) \textbf{Proposed w/o Style Decoder}:
The style decoder is removed, both the outputs $\mathcal{H}_{p}^{'}$ and $\mathcal{H}_{s}^{'}$ of the extended variance adaptor are fed to phoneme decoder together.
iii) \textbf{Proposed w/o Style Extractor}:
The style extractor is removed, which means the style loss $\mathcal{L}_{style}$ is removed during the TTS training stage.

CMOS is employed to compare the synthesized speech in terms of naturalness and expressiveness.
As shown in Table \ref{tab:ablation study},
the performance of the three settings 
is degraded to various degrees respectively compared with the proposed method.
This indicates that all these components have substantial impact on our proposed model.
Furthermore, the results also indicate that both our proposed style pre-training strategy and the novel TTS architecture with two encoder-decoder paths can alleviate the 
role and out-of-domain expressiveness deterioration problem caused by unbalanced style distribution and 
insufficient model generalizability.

\vspace{-10pt}
\begin{table}[h]
\vspace{-5pt}
\caption{CMOS comparison for ablation study.}
\label{tab:ablation study}
\centering
\begin{tabular}{l|c}
\toprule
\textbf{Model}              & \textbf{CMOS} \\ \hline
Proposed           & \textbf{0}    \\ 
\hspace{6mm} w/o Style Encoder    & -0.142      \\ 
\hspace{6mm} w/o Style Decoder   & -0.133     \\ 
\hspace{6mm} w/o Style Extractor    & -0.150     \\ 
\bottomrule
\end{tabular}
\vspace{-5pt}
\end{table}

\vspace{-10pt}
\section{Conclusion}
\vspace{-5pt}
This work addresses the problem of poor expressiveness in audiobook speech synthesis due to generalized model architecture and unbalanced style distribution in the training data. 
We propose a pre-trained VQ-VAE-based style extractor
and a novel TTS architecture with two encoder-decoder paths. 
Both objective and subjective experiments demonstrate the 
effective
performance of our proposed method
in terms of naturalness and expressiveness of the synthesized speech, especially for the role and out-of-domain scenarios.
\bibliographystyle{IEEEbib.bst}
\bibliography{refs.bib}
\end{document}